\documentstyle[12pt,epsfig,amsfonts,latexsym]{article}

\makeatletter

\@addtoreset{equation}{section} \makeatother

\newcommand{\be}{\begin{equation}}

\newcommand{\ee}{\end{equation}}

\newcommand{\ba}{\begin{eqnarray}}

\newcommand{\ea}{\end{eqnarray}}

\newcommand{\bea}{\begin{eqnarray*}}

\newcommand{\eea}{\end{eqnarray*}}

\newcommand{\gsim}{\raise.3ex\hbox{$>$\kern-.75em\lower1ex\hbox{$\sim$}}}

\newcommand{\lsim}{\raise.3ex\hbox{$<$\kern-.75em\lower1ex\hbox{$\sim$}}}

\begin{document}

\titlepage

\begin{flushright}
\today \\
hep-th/0407025 \\
\end{flushright}
\vskip 1cm
\begin{center}
{\large \bf The 4d effective action of 5d
gauged supergravity with boundaries}
\end{center}

\vspace*{5mm} \noindent

\centerline{Ph.~Brax\footnote{brax@spht.saclay.cea.fr}${}$,
N.~Chatillon\footnote{chatillon@spht.saclay.cea.fr}${}$}

\vskip 0.5cm \centerline{$$ \em Service de Physique Th\'eorique}
\centerline{\em CEA/DSM/SPhT, Unit\'e de recherche associ\'ee au
CNRS,} \centerline{\em CEA-Saclay F-91191 Gif/Yvette cedex,
France.} \vskip 0.5cm

\

\begin{center}
{\bf Abstract}
\end{center}

We consider gauged five dimensional supergravity with boundaries
and vector multiplets in the bulk. We analyse the zero modes of
the  BPS configurations preserving $N=1$ supergravity at low
energy. We find the 4d low energy effective action involving the
moduli associated to the BPS zero modes. In particular, we derive
the K\"ahler potential on the moduli space corresponding to the low
energy 4d $N=1$ effective action.
\newpage
\section{Introduction}

Supergravity in five dimensions provides the framework to build a
host of interesting extra--dimensional models. Two particularly
important examples are the compactified version of M--theory on a
Calabi--Yau manifold \cite{ovrut} and the supersymmetric Randall--Sundrum
model \cite{Altendorfer:2000rr,Falkowski:2000er,kallosh,offshell,flp2,
Brax:2001xf,susyRS_arbitrary_tensions,Brax:2002vs,Lalak:2002kx,twisted_sugra,
twist_warp_sugra,flp3,effective_detuned}. In both cases, the fifth dimension is an interval
corresponding to a $Z_2$ orbifold with two fixed points. The four
dimensional end--points of the interval are two branes where
matter can be confined. In particular, these branes have a tension
leading to a warped gravitational background in the bulk. Hence
the type of extra--dimensions present in these supergravity models
differs drastically from the usual flat embedding of branes in
Minkowski space.  The warping  of the fifth dimension has led to
some interesting phenomenological developments concerning the
hierarchy problem \cite{ADD,RS1} and the cosmological constant problem
\cite{selftuning}.

Such brane models have two typical regimes. In the high energy
regime above the brane tension, peculiar phenomena, such as
quadratic terms in the matter density contributing to the
Friedmann equation, appear \cite{branecosmo}. At low energy below the brane tension,
the physics can be described by a four dimensional effective
action obtained by integration over the extra--dimension. The
dynamics are encoded in the moduli corresponding to supersymmetric
flat directions of the five dimensional models. For BPS
configurations preserving one half of the original supersymmetry,
the low energy action is a 4d $N=1$ action entirely specified by
its K\"ahler potential expressed in terms of the moduli.

In the present paper, we will focus on supergravity in singular
spaces as formulated by Bergshoeff, Kallosh and Van Proeyen \cite{kallosh}.
It is
defined as gauged supergravity in 5d with $n$ vector multiplets in
the bulk coupled to two boundary branes. The moduli space is
$2(n+1)$ dimensional. In section 2 we recall some properties of
the moduli space both for the real moduli and the associated axion
fields. In section 3, we analyse the low energy effective action
and give a closed expression for the K\"ahler potential in terms of
the moduli.
\section{Supergravity with Boundary Branes}
The bulk theory is $N=2$ pure supergravity \cite{5d pure sugra} coupled to arbitrary vector multiplets \cite{vector coupling}. We will not treat the most general case with hypermultiplets and tensor multiplets, which were coupled in \cite{vector-tensor-hyper}.
Gauged supergravity with boundary branes in five dimensions has
been elegantly constructed when vector multiplets live in the bulk
\cite{kallosh,offshell}. The supergravity multiplet comprises the
metric tensor $g_{ab}$, $a,b=1\dots 5$, the gravitini $\psi_a^A$
where $A=1,2$ is an $SU(2)_R$ index and the graviphoton field
$A_a$. The $N$=2 vector multiplets in the bulk possess one vector
field, a $SU(2)_R$ doublet of symplectic Majorana spinors and one
real scalar. When considering $n$ vectors multiplets, it is
convenient to denote by $A^I_a$, $ I=1\dots n+1$, the $(n+1)$
vector fields including the graviphoton.

The vector multiplets comprise scalar fields $\phi^i$
parametrising the manifold $M$
\begin{equation}
C_{IJK}h^I(\phi) h^J(\phi) h^K(\phi)=1
\end{equation}
with the functions $h^I(\phi), I=1\dots n+1$ playing the role of
auxiliary variables.  The manifold $M$ has dimension $n$. Defining
the metric
\begin{equation}
G_{IJ}\equiv -2C_{IJK}h^{K}+3h_Ih_J
\end{equation}
where $h_I\equiv C_{IJK}h^Jh^K$, the bosonic part of the Lagrangian
(vector fields not included) reads
\begin{equation}
S_{bulk}=\frac{1}{2\kappa_5^2}\int d^5x \sqrt {-g_5}\Big({\mathcal
R} -\frac{3}{4}(g_{ij}\partial_{\mu}
\phi^i\partial^{\mu}\phi^j+V)\Big) \label{lag}
\end{equation}
where the sigma-model metric $g_{ij}$ is
\begin{equation}
g_{ij}=2 G_{IJ}\frac{\partial h^I}{\partial \phi^i}\frac{\partial
h^J}{\partial \phi^j}
\end{equation}
and the potential is given by
\begin{equation}
V=U_iU^i-U^2
\end{equation}
where $U_i=\frac{\partial U}{\partial\phi^i}$ and indices are
raised  using the sigma-model metric $g^{ij}$. Notice that the
metric can be written as
\begin{equation}
G_{IJ}= -\frac{1}{3} \frac{\partial^2}{\partial h^I\partial h^J}
\ln (C_{PQR} h^P h^Q h^R)\vert_M
\end{equation}
where the constraint defining $M$ is only used after the two
derivatives have been computed.
 The superpotential $U$ defines the dynamics of the
theory. It is given by
\begin{equation}
U=4\sqrt\frac{2}{3}g h^Iq_I
\end{equation}
where $g$ is a gauge coupling constant and the $q_I$'s are real
numbers such that the $U(1)$ gauge field is $A^I_a q_I$.

The boundary action depends on two  fields. There is a
supersymmetry singlet  $G$ and a four form $A_{\mu\nu\rho\sigma}$
\cite{kallosh}. One also modifies  the bulk action by replacing
$g\to G$ and adding  a direct coupling
\begin{equation}
S_A=\frac{2}{4! \kappa_5^2}\int d^5x
\epsilon^{abcde}A_{abcd}\partial_{e}G.
\end{equation}
The boundary action is taken as
\begin{equation}
S_{bound}=-\frac{1}{\kappa_5^2}\int d^5x
(\delta_{x_5}-\delta_{x_5-R}) (\sqrt {-g_4}\frac{3}{2}U
+\frac{2g}{4!}\epsilon^{\mu\nu\rho\sigma}A_{\mu\nu\rho\sigma}).
\label{boundary action}
\end{equation}
where $\mu,\nu,\rho,\sigma$ are four-dimensional indices on the
branes. Notice that the four-form $A_{abcd}$ is not dynamical.

The supersymmetry algebra closes on shell where
\begin{equation}
G(x)=g\epsilon (x_5),
\end{equation}
and $\epsilon (x_5)$ jumps from -1 to 1 at the origin of the fifth
dimension. On shell the bosonic Lagrangian reduces to  the bulk
Lagrangian  coupled to the boundaries as,
\begin{equation}
S_{bound}=-\frac{3}{2\kappa_5^2}\int d^5x
(\delta_{x_5}-\delta_{x_5-R})\sqrt {-g_4}U,
\end{equation}
Crucially, the boundary branes couple directly to the bulk
superpotential. Notice that the two branes have opposite
(field-dependent) tensions
\begin{equation}
\lambda_{\pm}=\pm \frac{3}{2\kappa_5^2}U
\end{equation}
where the first brane has positive tension.

The gauge fields in the bulk have a kinetic term parametrised by
the metric $G_{IJ}$
\begin{equation}
S_{gauge}= -\frac{1}{4\kappa_5^2}\int d^5 x \sqrt{-g_5} G_{IJ}
F^I_{ab}F^{Jab}
\end{equation}
The dimensional reduction of this term to 4d will lead to the
axion fields at low energy.

We will study BPS configurations preserving one half of
supersymmetry. These BPS configurations are associated to a
complex $(n+1)$--dimensional moduli space with a K\"ahlerian structure. The low
energy action of 5d supergravity with boundaries reduces to a
supergravity theory in 4d whose structure depends only on the
K\"ahler potential on the moduli space. The moduli space comprises
$(n+1)$ real directions associated to $n$ scalar fields
corresponding to tangent directions to the manifold $M$ and one
scalar mode of gravitational origin, the radion, which can either
be seen as the distance between the branes or the $55$ component
of the bulk metric. On the moduli space, there is no potential and
therefore no superpotential as the moduli correspond to
supersymmetric flat directions.

\section{The Moduli Space}

The previous theory admits flat directions where the original
supersymmetry is broken to $N=1$ in 4d. These BPS configurations
are obtained by requiring that the gravitino and gaugino
variations vanish. The BPS backgrounds are determined by first
order differential equations for which the boundary conditions at
the branes are automatically satisfied. The moduli space of the
theory is parametrised by the constants of integrations of the BPS
equations, corresponding to BPS zero modes. For $n$ vector
multiplets, there are $(n+1)$ moduli. These moduli are associated
to as many axion fields, the whole moduli space becoming a K\"ahler
manifold whose K\"ahler potential will be determined later.

The BPS equations corresponding to the presence of Killing spinors
are given by \cite{kallosh}
\begin{equation}
\frac{d\phi^x}{dz}=g^{xy}\frac{\partial W}{\partial \phi^y}, \ \frac{d\ln \tilde a }{dz}=-\frac{W}{4}
\end{equation}
where the bulk metric has been written
\begin{equation}
ds^2=dz^2 + \tilde a^2(z) \eta_{\mu\nu}dx^\mu dx^\nu
\end{equation}
 It is convenient to define
\begin{equation}
dy= \tilde a^2(z) dz
\end{equation}
so that the metric (with $a(y)=\tilde a(z)$)
\begin{equation}
ds^2=\frac{dy^2}{a^4(y)} + a^2(y) \eta_{\mu\nu}dx^\mu dx^\nu
\end{equation}
yields an effective action in the Einstein frame when integrating over the fifth dimension. From this one gets that
\begin{equation}
g_{xy}\frac{d\phi^y}{dy}=2G_{IJ}\frac{\partial h^I}{\partial \phi^x}\frac{\partial h^J}{\partial y}
\end{equation}
Using the relation $\frac{\partial h_I}{\partial y}=-G_{IJ}\frac{\partial h^J}{\partial y}$, we find that
\begin{equation}
h^I_x(\frac{d(a^2 h_I)}{dy}+\frac{\epsilon (y)}{2} q_I)=0
\end{equation}
where we have used $h^I_xh_I=0$.
Similarly we get that
\begin{equation}
h^I( \frac{d(a^2h_I)}{dy} +\frac{\epsilon (y)}{2} q_I)=0
\end{equation}
Notice that the $h^I_x$'s form a basis of vectors corresponding to
the tangent space $TM$  and the normal space to $M$ is
parametrised by $h^I$. Being orthogonal to a basis of the
$(n+1)$--dimensional space where the moduli space is embedded, we
deduce that
\begin{equation}
\frac{d(a^2h_I)}{dy} +\frac{\epsilon (y)}{2} q_I=0
\end{equation}
leading to
\begin{equation}
\tilde h_I\equiv a^2 h_I = t_I - \frac{1}{2} q_I \vert y\vert
\end{equation}
where $t_I$ is an integration constant. There are thus $(n+1)$
integration constants. These $(n+1)$ integration constants
parametrise the real part of the moduli space of the theory.

The scale factor can be obtained via
\begin{equation}
a^2 = \tilde h_I h^I
\end{equation}
or equivalently
\begin{equation}
a^3= C_{IJK} \tilde h^I \tilde h^J \tilde h^K \label{constraint}
\end{equation}
where $\tilde h ^I = a h^I$. These variables are solutions of
\begin{equation}
\tilde h_I = C_{IJK} \tilde h^J \tilde h^K
\end{equation}
in such a way that $ \tilde h^I$ is a function of $\tilde h_I$. Note that according to (\ref{constraint}), the (n+1) variables $h^I$ defined as $\tilde h^I(\tilde h_K)/a(\tilde h_K)$ automatically belong to the n-dimensional  manifold $M$.

Let us now find the imaginary parts associated to the real moduli.
These axion fields are associated with the fifth components of the
bulk gauge fields. Let us assume that $A^I_5$ is the only
non-vanishing component ; equivalently, we choose the $A^I_5$ field even under the orbifold parity, while $A^I_{\mu}$ is odd. The effective theory will consequently contain no $N=1$ vector multiplet. The equations of motion read
\begin{equation}
\partial_a(\sqrt{-g} g^{ac}g^{bd} G_{IJ} F^{Jcd})=0
\end{equation}
We find that for $a=5$
\begin{equation}
 F^I_{\mu 5}=\frac{1}{a^4}  G^{IJ}\partial_\mu b_J (x)
\end{equation}
The $a=\mu$ case reduces to
\begin{equation}
\Box^{(4)}b_I=0
\end{equation}
Hence we find $(n+1)$ axion zero modes.

On the whole the moduli space comprises $2(n+1)$ scalar fields
$t_I$ and $b_I$.

\section{The K\"ahler Potential}

At low energy the bulk metric can be parametrised using
\begin{equation}
ds^2= \frac{dy^2}{a^4(y,x)} + a^2(y,x) g_{\mu\nu}(x) dx^\mu dx^\nu
\end{equation}
where $g_{\mu\nu}$ is a metric corresponding to the graviton zero
mode, and the warp factor depends on all coordinates implicitly
through  $\tilde h_I(x,y)=t_I(x)-\frac{1}{2}q_I y$. Note that the
branes are straight in this coordinate system. It leads to 4d
gravity at low energy as the 5d Einstein--Hilbert term leads to
\begin{equation}
\frac{1}{2\kappa_5^2}\int d^4x dy \sqrt{-g_5} {\cal R}\supset
\frac{1}{2\kappa_4^2} \int d^4 x \sqrt{-g} {\cal R}^{(4)}
\end{equation}
where ${\cal R}^{(4)}$ is the Ricci scalar associated to $g_{\mu\nu}$ and
\begin{equation}
\frac{1}{\kappa_4^2}=\frac{2d}{\kappa_5^2}
\end{equation}
with $d= \int_{y_+}^{y_-} dy=\frac{1}{2}\int dy$ evaluated between the two branes
located at constant $y= y_{\pm}$.  Hence the ansatz leads to 4d gravity in
the Einstein frame. We will see later that the implicit dependence of
$a(y,x)$ on $x$ due to the dependence on $t_I(x)$ leads to a
contribution to the moduli kinetic terms.

 We now
promote the axions to be fields $b_I(x)$ and compute the kinetic
terms resulting from the gauge field kinetic term in the
Lagrangian. We obtain that the axions have  kinetic terms
\begin{equation}
S_{axion}= -\frac{1}{4\kappa_5^2}\int d^4x \sqrt{-g}(\int dy
\frac{1}{a^4} G^{IJ})g^{\mu\nu}\partial_\mu b_I  \partial_\nu b_J
\end{equation}
It has the form of a non-linear sigma model with a metric
\begin{equation}
K^{I\bar J}= \frac{1}{\kappa_5^2}\int dy \frac{1}{a^4} G^{IJ}
\end{equation}
We will see later that this metric derives from a K\"ahler
potential.

The kinetic terms for the real moduli follow from
\begin{equation}
g_{xy}\partial_\mu \phi^x \partial^\mu \phi^y =
\frac{(G^{IJ}-h^Ih^J)}{a^4}g^{\mu\nu}\partial_\mu t_I \partial
_\nu t_J \label{kin}
\end{equation}
where we have used the fact that $\tilde h_I$ depends on $x$ only
via $t_I$. Notice too that we have used the fact that the kinetic
terms coming from the scalar fields are such that
\begin{equation}
(G^{IJ}-h^Ih^J)\partial_\mu h_I \partial ^\mu
h_J=G^{IJ}\partial_\mu h_I \partial ^\mu h_J
\end{equation}
 corresponding to the
projection of the metric $G^{IJ}$ to the tangent space of $M$.
Hence the kinetic terms coming from the scalar fields in 5d only
involve $n$ moduli.  Note also that
\begin{equation}
(G^{IJ}-h^Ih^J)\partial_\mu h_I=
\frac{(G^{IJ}-h^Ih^J)}{a^2}\partial_\mu  t_I
\end{equation}
and therefore the previous result (\ref{kin}).

 The
Einstein-Hilbert term in 5d can be evaluated and leads to a
contribution to the moduli kinetic terms
\begin{equation}
\int d^5 x \sqrt{-g_5} {\cal R}\supset  -\frac{3}{2}\int d^4 x dy
g^{\mu\nu} \frac{\partial_\mu a^2\partial_\nu a^2}{a^4}
\end{equation}
where
\begin{equation}
\partial_\mu a^2= h^I \partial_\mu t_I
\end{equation}
implying that the Einstein-Hilbert term contributes as
\begin{equation}
-\frac{3}{2}\int d^4 x dy g^{\mu\nu}h^Ih^J \frac{\partial_\mu t_I\partial_\nu t_J}{a^4}
\end{equation}
This term corresponds to a projection of the moduli kinetic terms
on the normal to $M$. As can be seen from its origin, i.e. the
dependence of the scale factor $a$ on $x$, it is associated to the
variations of the $55$ component of the bulk metric, i.e. the
radion.

Collecting the factor from the scalar field kinetic terms and the Einstein-Hilbert we obtain that the kinetic terms of the scalar fields lead to
\begin{equation}
-\frac{3}{4\kappa_5^2}\int d^4x \sqrt{-g}(\int dy
\frac{G^{IJ}}{a^4})g^{\mu\nu}\partial_\mu t_I  \partial_\nu t_J
\end{equation}
Defining the complex moduli
\begin{equation}
T_I= \sqrt{\frac{3}{2}} t_I +i \frac{b_I}{\sqrt 2}
\end{equation}
and using complex variable notations we find that the kinetic terms read
\begin{equation}
S_{moduli}=-\frac{1}{2\kappa_5^2}\int d^4x \sqrt{-g}(\int dy
\frac{G^{I J}}{a^4})g^{\mu\nu}\partial_\mu T_I
 \partial_\nu \bar T_J
\end{equation}

To complete our description of the moduli space, we need to show
that
\begin{equation}
K^{I\bar J}=\frac{1}{\kappa_5^2}\int dy \frac{G^{IJ}}{a^4}
\end{equation}
is a second derivative.
Let us define
\begin{equation}
F(t_I)=C^{IJK}\tilde h_I   \tilde h_J \tilde h_K
\end{equation}
where $C^{IJK}\equiv G^{IL}G^{JM}G^{KP}C_{LMP}$. Using the fact that
\begin{equation}
h_I h_J \partial C^{IJK}=0
\end{equation}
where the derivative $\partial$ is taken with respect to $\tilde h_I$,
one finds that
\begin{equation}
\frac{\partial^2 \ln F}{\partial \tilde h_I \partial \tilde h_J}=-
3 \frac{G^{IJ}}{a^4}
\end{equation}
We have used  $h_I \partial h^I=0$ and $G^{IJ}\partial h_J = - \partial h^I$ as $h^I(\tilde h_J)$
is always on $M$.
One obtains that
\begin{equation}
K^{I\bar J}=-\frac{1}{3\kappa_5^2}\frac{\partial^2}{\partial t_I \partial t_J}\int dy  \ln ( F)
\end{equation}
This is the expected result implying that the K\"ahler potential is
given by
\begin{equation}
K=-\frac{4}{3\kappa_5^2}\int dy \ln F(\frac{T_I+\bar T_{\bar
I}}{2})
\end{equation}
as a function of $t_I=(T_I+\bar T_{\bar I})/2$, or equivalently
\begin{equation}
K= -\frac{4}{3\kappa_5^2} \int dy \ln \Big[ C^{IJK}(\frac{(T_I+\bar T_{\bar I})}{2}-\frac{q_I}{2}y) (\frac{(T_J+\bar T_{\bar J})}{2}-\frac{q_J}{2}y)
(  \frac{(T_K+\bar T_{\bar K})}{2}-\frac{q_K}{2}y)\Big]
\end{equation}
which depends only on the real moduli $t_I$ only.

Let us illustrate this general result with two examples.
In the case of non-gauged supergravity the K\"ahler potential reads
($q_I=0$)
\begin{equation}
K= -\frac{8d}{3\kappa_5^2}\ln \Big( C^{IJK}\frac{T_I+\bar T_{\bar
I}}{2}\ \frac{T_J+\bar T_{\bar J}}{2}\
  \frac{T_K+\bar T_{\bar K}}{2}\Big)
\end{equation}
As the metric $G_{IJ}$ does not depend on $y$, we can define
\begin{equation}
{\cal T}^I= G^{IJ}T_J
\end{equation}
leading to
\begin{equation}
K= -\frac{8d}{3\kappa_5^2} \ln \Big( C_{IJK}\frac{{\cal T}^I+\bar {\cal T}^{\bar I}}{2} \ \frac{{\cal T}^J+\bar {\cal T}^{\bar J}}{2} \  \frac{{\cal T}^K+\bar
{\cal T}^{\bar K}}{2}\Big)
\end{equation}
where $d=\int_{y_+}^{y_-} dy=\frac{1}{2}\int dy$. It coincides with the known result
in linearised M-theory \cite{ovrut} and the recent analysis
\cite{ferrara}.
\\

The simplest  gauged case is given by a one vector multiplet model
($n=1$) with only non-zero coefficient $C_{112}=1$ (and
permutations thereof). The defining relation for the field
manifold M is then $3(h^1)^2h^2=1$. Solving the constraint
\begin{equation}
\tilde h_I = C_{IJK} \tilde h^J \tilde h^K
\end{equation}
we  find
\begin{equation}
\tilde h^1=\sqrt{\tilde h_2},\ \tilde h^2=\frac{\tilde h_1}{2\sqrt{\tilde h_2}}
\end{equation}
and  the warp factor
\begin{equation}
a(y,x)=(\frac{3}{2}\tilde h_1 \sqrt{\tilde h_2})^{1/3}
\end{equation}
The metric is now diagonal
\begin{eqnarray}
G_{11}&=&(\frac{2}{3})^{1/3}(\frac{\tilde h_1}{\tilde h_2})^{2/3}
\nonumber \\
G_{22}&=&2(\frac{2}{3})^{1/3}(\frac{\tilde h_2}{\tilde h_1})^{4/3}
\nonumber \\
G_{12}&=&0
\end{eqnarray}
and the only non-zero entry is
\begin{equation}
C^{112}=(G_{11})^{-2}(G_{22})^{-1}C_{112}=3/4
\end{equation}
and permutations thereof. The K\"ahler potential is finally
\begin{eqnarray}
K&=&-\frac{32d}{3q_1\kappa_5^2}\Big[(t_1-\frac{q_1}{2}y_+)\ln(t_1-\frac{q_1}{2}y_+)-(t_1-\frac{q_1}{2}y_-)\ln(t_1-\frac{q_1}{2}y_-)\Big]
\nonumber \\
&&-\frac{16d}{3q_2\kappa_5^2}\Big[(t_2-\frac{q_2}{2}y_+)\ln(t_2-\frac{q_2}{2}y_+)-(t_2-\frac{q_2}{2}y_-)\ln(t_2-\frac{q_2}{2}y_-)\Big]
\nonumber \\
&&+\frac{16d}{3\kappa_5^2}(\frac{3}{2}-\ln\frac{3}{2})
\end{eqnarray}
such that the K\"ahler metric $K^{I\bar J}=\frac{\partial^2 K}{\partial T_I\partial\bar T_J}$ is diagonal
\begin{equation}
K^{1\bar 1}= \frac{4d^2}{3\kappa_5^2}\frac{1}{(t_1 -
\frac{q_1}{2}y_+)(t_1 - \frac{q_1}{2}y_-)},\ K^{2\bar 2}=
\frac{2d^2}{3\kappa_5^2}\frac{1}{(t_2 - \frac{q_2}{2}y_+)(t_2-
\frac{q_2}{2}y_-)}
\end{equation}
In the ungauged case $q_I=0$ this simplifies to
\begin{equation}
K=-\frac{16d}{3\kappa_4^2}\ln(\frac{T_1+\bar
T_1}{2})-\frac{8d}{3\kappa_4^2}\ln(\frac{T_2+\bar
T_2}{2})-\frac{16d}{3\kappa_4^2} \ln\frac{3}{2}
\end{equation}

 Let us now introduce matter on
the boundary branes. We couple the matter fields to the induced
metric on the ith--brane leading to an action for the matter
scalar field $s$ coupled to the moduli
\begin{equation}
\int d^4 x \sqrt{-g} \Big( a^2(t_I) (\partial s\partial \bar s)+
a^4(t_I) \vert\frac{\partial w(s)}{\partial s}\vert ^2 \Big)
\end{equation}
up to derivative terms in the $t_I$'s. We have denoted by $w$ the
superpotential of the supersymmetric theory on the brane. As we
are supersymmetrising the matter action only at zeroth order in
$\kappa_4$, we have suppressed the non-renormalizable terms in the
matter fields for fixed moduli, hence the globally  supersymmetric
form of the potential. Such an action can be supersymmetrised
\begin{equation}
-\int d^4 x d^4 \theta E^{-1}  a^2(\frac{T_I +\bar T_{\bar
I}}{2})\Sigma \bar \Sigma
\end{equation}
where $\Sigma =s+\dots $ is the chiral superfield of matter on the
brane. Similarly the potential on the brane follows from
\begin{equation}
\int d^4x d^2\theta \Phi^3 W(T_I,\Sigma)
\end{equation}
where
\begin{equation}
W(T_I,S)= a^3(T_I)w(\Sigma)
\end{equation}
and $\Phi$ is the chiral compensator whose $F$--term is the
gravitational scalar auxiliary field.  At low energy this leads to a
direct coupling between matter fields and the moduli. When matter
is on both branes, a sum over the branes contributions evaluated
at the brane positions is understood.

A particularly interesting case corresponds to constant superpotentials on the branes
as a function of the moduli. In that case, the tensions of the branes are shifted from their BPS values.
When only the superpotential on the second brane does not vanish, this is a hidden brane scenario
of supersymmetry breaking, as was studied in \cite{soft} in the case of a single bulk vector multiplet. The analysis of the corresponding physics is left for future work.

\section{Conclusion}

We have studied 5d gauged supergravity with an arbitrary number of
vector multiplets and boundaries. In particular we have focused on
the low energy effective action parametrised by the moduli of BPS
configurations preserving $N=1$ supersymmetry in 4d. The 4d
effective action is determined by the K\"ahler potential expressed
in terms of the moduli. We have given a closed expression for the
K\"ahler potential in terms of the warping of the 5d metric and
the cubic polynomial defining the real--special geometry of 5d
supergravity with vector multiplets.

The coupling of the moduli to matter on the branes has also been
made explicit. In particular, this may be useful in analysing the
way supersymmetry breaking may be generated in 5d brane models.


\begin{thebibliography}{99}

\bibitem{ovrut}
A. Lukas, B. A. Ovrut, K.S. Stelle and D. Waldram, {\it Heterotic
M-theory in Five Dimensions}, Nucl. Phys. B {\bf 552} (1999) 246
[arXiv:hep-th/9806051].

\bibitem{Altendorfer:2000rr}
R.~Altendorfer, J.~Bagger and D.~Nemeschansky, {\it Supersymmetric
Randall-Sundrum scenario}, Phys.\ Rev. D {\bf 63} (2001) 125025
[arXiv:hep-th/0003117].

\bibitem{Falkowski:2000er}
A.~Falkowski, Z.~Lalak and S.~Pokorski, {\it Supersymmetrizing
branes with bulk in five-dimensional supergravity}, Phys.\ Lett.\
B {\bf 491} (2000) 172 [arXiv:hep-th/0004093].

\bibitem{kallosh}
E.~Bergshoeff, R.~Kallosh and A.~Van Proeyen, {\it Supersymmetry
in singular spaces}, JHEP {\bf 0010} (2000) 033
[arXiv:hep-th/0007044].

\bibitem{offshell}
M. Zucker, {\it Supersymmetric brane world scenarios from
off-shell supergravity}, Phys. Rev. D {\bf 64} (2001) 024024
[arXiv:hep-th/0009083].

\bibitem{flp2}
A. Falkowski, Z. Lalak and S. Pokorski, {\it Five-dimensional
gauged supergravities with universal hypermultiplet and warped
brane worlds}, Phys. Lett. B {\bf 509} (2001) 337
[arXiv:hep-th/0009167].

\bibitem{Brax:2001xf}
Ph.~Brax, A.~Falkowski and Z.~Lalak, {\it Non-BPS branes of
supersymmetric brane worlds}, Phys.\ Lett.\ B {\bf 521} (2001) 105
[arXiv:hep-th/0107257].

\bibitem{susyRS_arbitrary_tensions}
J. Bagger and D. Belyaev, {\it Supersymmetric branes with (almost)
arbitrary tensions}, Phys. Rev. D {\bf 67} (2003) 025004
[arXiv:hep-th/0206024].

\bibitem{Brax:2002vs}
P.~Brax and Z.~Lalak, {\it Brane world supersymmetry, detuning,
flipping and orbifolding}, Acta Phys.\ Polon.\ B {\bf 33} (2002)
2399 [arXiv:hep-th/0207102].

\bibitem{Lalak:2002kx}
Z.~Lalak and R.~Matyszkiewicz, {\it On Scherk-Schwarz mechanism in
gauged five-dimensional supergravity and  on its relation to
bigravity}, Nucl.\ Phys.\ B {\bf 649} (2003) 389
[arXiv:hep-th/0210053].

\bibitem{twisted_sugra}
Z. Lalak and R. Matyszkiewicz, {\it Twisted supergravity and
untwisted super-bigravity}, Phys.Lett. B {\bf 562} (2003) 347-357
[arXiv:hep-th/0303227].

\bibitem{twist_warp_sugra}
J. Bagger and D. Belyaev, {\it Twisting Warped Supergravity}, JHEP
{\bf 0306} (2003) 013 [arXiv:hep-th/0306063].

\bibitem{flp3}
A.~Falkowski, Z.~Lalak and S.~Pokorski, {\em Four dimensional
supergravities from five dimensional brane worlds}, Nucl. Phys. B
{\bf 613} (2001) 189-217 [arXiv:hep-th/0102145].

\bibitem{effective_detuned}
J. Bagger and M. Redi, {\it Radion effective theory in the detuned
Randall-Sundrum model}, JHEP {\bf 0404} (2004) 031
[arXiv:hep-th/0312220].

\bibitem{ADD}
N. Arkani-Hamed, S. Dimopoulos and G. Dvali, {\it The Hierarchy Problem
and New Dimensions at a Millimeter}, Phys.Lett. B{\bf 429} (1998) 263-272
[arXiv:hep-ph/9803315].

\bibitem{RS1}
L. Randall and R. Sundrum, {\it A large mass hierarchy from a
small extra dimension}, Phys. Rev. Lett. {\bf 83} (1999) 3370
[arXiv:hep-ph/9905221].

\bibitem{selftuning}
N. Arkani-Hamed, S. Dimopoulos, N. Kaloper and R. Sundrum, {\it A
Small Cosmological Constant from a Large Extra Dimension},
Phys.Lett. B{\bf 480} (2000) 193-199 [hep-th/0001197] ; S. Kachru,
M. Schulz and E. Silverstein, {\it Self-tuning flat domain walls
in 5d gravity and string theory}, Phys.Rev. D{\bf 62} (2000)
045021 [arXiv:hep-th/0001206].

\bibitem{branecosmo}
P. Binetruy, C. Deffayet, U. Ellwanger and D. Langlois, {\it Brane cosmological evolution in a bulk with cosmological constant}, Phys. Lett. B477 (2000) 285-29 [arXiv:hep-th/9910219].

\bibitem{5d pure sugra}
E. Cremmer, {\it Supergravities in 5 dimensions}, in {\it Superspace and
Supergravity}, Eds. S.W. Hawking and M. Rocek (Cambridge univ. press,
1981), 267.

\bibitem{vector coupling}
M. Gunaydin, G. Sierra and P.K. Townsend, {\it The geometry of N=2
Einstein supergravity and Jordan algebras}, Nucl. Phys. B{\bf 242} (1984) 244 ;
{\it Gauging the d=5 Maxwell-Einstein supergravity theories : more on Jordan
algebras}, Nucl. Phys. B{\bf 253} (1985) 573.

\bibitem{vector-tensor-hyper}
A. Ceresole and G. Dall'Agata, {\it General matter coupled N=2, D=5 gauged
supergravity}, Nucl. Phys. B{\bf 585} (2000) 143-170 [arXiv:hep-th/0004111].

\bibitem{ferrara} L. Andrianopoli, S. Ferrara and M. A. Lled\'o,
{\it Scherk--Schwarz reduction of $D=5$ special and quaternionic
geometry} [arXiv:hep-th/0405164].

\bibitem{soft}
Ph. Brax and N. Chatillon, {\it Soft supersymmetry breaking on the brane} [arXiv:hep-th/0405143].

\end{thebibliography}
\end{document}